\documentclass[10pt]{iopart}

\usepackage{amssymb}
\usepackage{graphicx}
\usepackage{epsfig}
\usepackage{indentfirst}
\usepackage{cite}
\usepackage{subfig}
\usepackage{cleveref}

\begin{document}

%\title[Fluctuation-induced switching of superconducting
%current-carrying bridges]{Fluctuation-induced switching of
%superconducting current-carrying bridge to a resistive state}

\title[Threshold fluctuations in superconducting
current-carrying bridge]{Threshold fluctuations in superconducting
current-carrying bridge}

\author{P M Marychev and D Yu Vodolazov}

\address{Institute for Physics of Microstructures, Russian Academy of Sciences, Nizhny Novgorod, 603950 Russia}

\ead{observermp@yandex.ru}

\begin{abstract}
We calculate the energy of threshold fluctuation $\delta F_{thr}$
which triggers the transition of superconducting current-carrying
bridge to resistive state. We show that the dependence $\delta
F_{thr}(I)\propto I_{dep}\hbar(1-I/I_{dep})^{5/4}/e$, found by
Langer and Ambegaokar for a long bridge with length $L \gg \xi$,
holds far below the critical temperature both in dirty and clean
limits (here $I_{dep}$ is the depairing current of the bridge and
$\xi$ is a coherence length). We also find that even 'weak' local
defect (leading to the small suppression of the critical current
of the bridge $I_c \lesssim I_{dep}$) provides $\delta
F_{thr}\propto I_c\hbar(1-I/I_c)^{3/2}/e$, typical for a short
bridge with $L \ll \xi$ or a Josephson junction.
\end{abstract}


%\submitto{\SUST}

\maketitle

\ioptwocol

\section{Introduction}

Superconducting state of bridge or wire with current is stable
with respect of infinitesimally small perturbations of
superconducting order parameter $\Delta$ up to currents close to
critical (depairing) current. But, if fluctuation induced change
of $\Delta$ is sufficiently large, instability is developed in the
superconductor even at $I<I_c$, leading to the appearance of a
finite resistance and dissipation. Theoretically,
fluctuation--induced switching first was studied in the work of
Langer and Ambegaokar (LA)~\cite{PhysRev-164}. They considered
long (length $L \gg \xi$, $\xi$ --- is the coherence length)
quasi-one-dimensional (transverse dimensions smaller than $\xi$)
superconducting bridge. To calculate threshold fluctuation LA
proposed to find a saddle-point state in the system nearest in
energy to the ground state. In their work authors obtained that
threshold fluctuation corresponds to a partial suppression of the
superconducting order parameter in a finite segment of the bridge
with size of about $\xi$ and derived the dependence of the energy
of threshold fluctuation on the applied current. Their result is
described well by following approximate expression~\cite{PRB-2003}
\begin{eqnarray}
 \label{df-LA}
\delta F_{LA}=\frac{4\sqrt{2}}{3}F_0 \left
(1-\frac{I}{I_{dep}}\right)^{5/4}
\\
 \nonumber
=\frac{\sqrt{6}}{2}\frac{I_{dep}\hbar}{e}\left
(1-\frac{I}{I_{dep}}\right)^{5/4},
\end{eqnarray}
where $F_0=\Phi_0^2S/32\pi^3\lambda^2\xi$, $\Phi_0$ is the
magnetic flux quantum, $S=wd$ is the area of the cross section of
the bridge with the width $w$ and thickness $d$, $\lambda$ is the
London penetration depth of the magnetic field, and
$I_{dep}=2I_0/3\sqrt{3}$ ($I_0=c\Phi_0S/8\pi^2\lambda^2\xi$) is
the depairing current in the Ginzburg–-Landau model, which
coincides with the expected critical current of the long ($L \gg
\xi$) bridge. In the work~\cite{JETPL-2016} LA approach was
generalized for superconducting bridges with arbitrary length and
it was shown that dependence $\delta F_{thr}(I)$ tends to the
expression $\delta F_{thr}=\hbar I_c (1-I/I_c)^{3/2}/e$ for short
bridges ($L\ll\xi$, $I_c\propto 1/L$ is a critical current of the
bridge). This dependence is typical for the energy of threshold
fluctuation for Josephson junctions with a sinusoidal
current--phase relation~\cite{PRB-1974}.

The energy of threshold fluctuation also was calculated for long
bridge using microscopic approach~\cite{PRL-2007,JETPL-2010}. In
the work ~\cite{PRL-2007} temperature and current dependencies of
$\delta F_{thr}$ were calculated on the basis of the Eilenberger
equations~\cite{Eilenberger} for clean long superconducting bridge
with only one conducting channel. However, $\delta F_{thr}$ has
been significantly overestimated at finite current since the
contribution to $\delta F_{thr}$ owing to the work performed by
the current source was not taken into account. In the present work
on the basis of the Eilenberger equations we recalculate the
dependence $\delta F_{thr}(I)$ at different temperatures and find
the agreement with power--$5/4$ law up to $T=0.5T_c$ which
coincides with the result found in work \cite{JETPL-2010} with
help of Usadel equations \cite{Usadel} for long dirty bridge. We
argue that the relation $\delta F_{thr}(0) \sim I_{dep} \hbar/e$
found in framework of GL model (see Eq. (1)) approximately holds
in a broad temperature range below $T_c$ not only for long bridges
(dirty or clean ones) but for short bridges too, with the
replacement of $I_{dep}(T)$ by actual critical current of the
bridge $I_c(T)$.

Our interest to the role of defects on $\delta F_{thr}(I)$ is
motivated by recent experimental works
~\cite{Nature-2009,PRL-2011,Aref}. In the experiment one usually
measures many times the switching current $I_{sw}$ (which has
random value due to fluctuations) to find the average value
$\langle I_{sw} \rangle$ and the dispersion $\sigma$ which are
directly related to $\delta F_{thr}(I)$ (for explicit relation
between $\langle I_{sw} \rangle$, $\sigma$ and $\delta F_{thr}(I)$
see for example (2,3) in ~\cite{Aref}). Note, that alternatively
$F_{thr}(0)$ could be found from temperature dependence of
resistivity near $T_c$, because $R(T)\propto\exp(-\delta
F_{thr}(0)/kT)$~\cite{Tinkham_book}. Although in
~\cite{PRL-2011,Aref} experiments were done for long
superconducting bridges in wide temperature interval below $T_c$
the good agreement with power--$3/2$ law was found. To explain
this result Khlebnikov~\cite{Khlebnikov} recently has developed
the model which considered the bridge as a discrete set of nodes
connected by superconducting links and in his model he neglected
local suppression of the superconducting order parameter. Below we
show that power--$3/2$ law can be obtained for a long bridge in
the framework of LA model, if one takes into consideration
presence of defects in the bridge such as constrictions or local
variation of the critical temperature or mean path length. We
argue that dependence $\delta F_{thr}(I)$ can deviate from
power--$5/4$ law even in case of relatively "weak" defects, when
the critical current $I_c$ of the bridge with defect is not far
from the depairing current $I_{dep}$.

\section{Effect of defects on $\delta F_{thr}(I)$}

Here we consider a model system consisting of the superconducting
bridge with cross section $S$ and length $L$, which connects two
superconducting banks whose cross section has the area $S_{pad}\gg
S$. Assuming that the maximal characteristic transverse size
$d\sim\sqrt{S}\ll\xi$, the problem can be considered as
one-dimensional and only the dependence on the longitudinal
coordinate x is taken into account.

To consider the effect of defects on the dependence $\delta
F_{thr}(I)$, we use the Ginzburg-Landau theory. To determine the
energy of threshold fluctuation it is necessary to find the saddle
state of the system corresponding to the local maximum of the free
energy in presence of external current source. Since it is
stationary state (albeit unstable), it is described by the
Ginzburg--Landau equation
\begin{equation}
 \label{GL}
   \xi^2_{GL}(0)\nabla^2\Delta+(1-T/T_c-|\Delta|^2/\Delta^2_{GL}(0))\Delta=0,
\end{equation}
where $\xi_{GL}(0)$ and $\Delta_{GL}(0)$ are the coherence length
and the superconducting order parameter in the GL model at zero
temperature respectively \cite{Tinkham_book}.

We seek the solution in the form $\Delta(x)/\Delta_{GL}=
f(x)exp(i\varphi(x))$. Then the dimensionless Ginzburg--Landau
equation has the form
\begin{equation}
 \label{one-dim-GL}
   \frac{d^2f}{dx^2}-\frac{j^2}{f^3}+f-f^3=0,
\end{equation}
where the condition of the constant current in the system,
$I=const$, is used (here $j=f^2d\varphi/dx=I/S$ is the current
density in the bridge). In ~\eref{one-dim-GL} the magnitude of the
superconducting order parameter $f$, length, and current density
are measured in units of $\Delta_{GL}=\Delta_{GL}(0)\sqrt{1-t}$,
$\xi=\xi_{GL}(0)/\sqrt{1-t}$ and $j_0=I_0/S$ ($t=T/T_c$ is the
dimensionless temperature). \eref{one-dim-GL} should be
supplemented with boundary conditions at the ends of the bridge
\begin{equation}
        \label{bridge-bound}
        f\big|_{-\frac{L}{2}}=f\big|_{\frac{L}{2}}=1,
\end{equation}
which follow from the assumption about nearly zero current density
at banks, and thus the order parameter reaches its equilibrium
value $f=1$.

The energy of threshold fluctuation can be found using the
expression
\begin{equation}
 \label{fluct-energy}
\delta
F_{thr}=F_{saddle}-F_{ground}-\frac{\hbar}{2e}I\delta\varphi,
\end{equation}
where $\delta\varphi$ is the additional phase difference between
the ends of the bridge appearing in the saddle-point state and
$F_{saddle}$ and $F_{ground}$ are the free energies of the
saddle-point and ground states, respectively. In our units these
energies take the form
\begin{equation}
 \label{free-energy}
   F_{saddle,ground}=-\frac{F_0}{2}\int f^4
   dx.
\end{equation}

Equation~\eref{one-dim-GL} with boundary conditions
~\eref{bridge-bound} is solved numerically for bridge with length
$L=30\xi$. In the numerical solution, we use the relaxation
method: the time derivative $\partial f/\partial t$ is added to GL
equation ~\eref{one-dim-GL} and iterations are performed until the
time derivative become zero within a specified accuracy. To find
the saddle- point state, we use the numerical method proposed in
\cite{Vodolazov_PRB}: at a given current, we fix the magnitude of
the order parameter $f(0)$ at the center of the bridge and allow
$f$ to change at all other points. The state with the minimum
fixed $f(0)$ value for which a steady-state solution exists is a
saddle-point state.

We consider three types of defects. The first type corresponds to
variation of critical temperature $T_c$ along the bridge. To
describe such defect in the model, we write  GL equation at the
defect region (placed in the center of the bridge) in the form
\begin{equation}
 \label{one-dim-GL-def}
   \frac{d^2f}{dx^2}-\frac{j^2}{f^3}+\alpha f-f^3=0,
\end{equation}
where the parameter $\alpha=(1-t^{\ast})/(1-t)$ characterizes the
deviation from the critical temperature of the rest of the bridge
(here $t^{\ast}=T/T^{\ast}_c$). Absence of a defect corresponds to
the case $\alpha=1$, and decrease of local critical temperature
$T^{\ast}_c<T_c$ corresponds to $\alpha<1$.

\begin{figure}[hbt]
 \begin{center}
\includegraphics[width=1.0\linewidth]{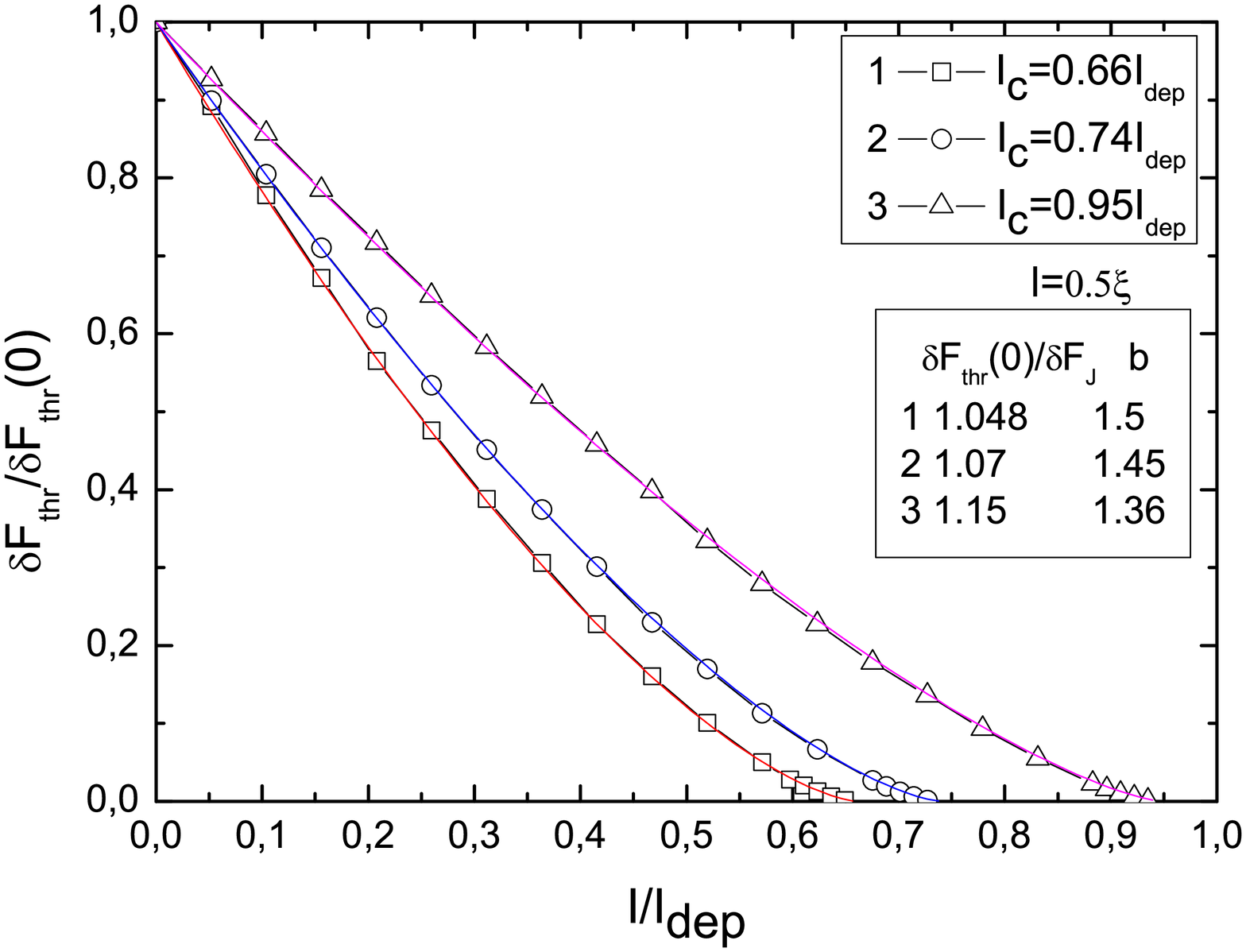}
 \caption{\label{Fig:vartc}
Dependence of the energy of threshold fluctuation on current for
bridges with local variation of $T_c$ (on length $l=0.5\xi$ in the
center of the bridge). Fitting functions $\delta
F_{thr}(0)(1-I/I_c)^b$ are shown by the solid lines, the
parameters $\delta F_{thr}(0)$ and $b$ are shown in inset. Here
$\delta F_J=\hbar I_c/e$.}
 \end{center}
\end{figure}

We consider defects with lengths $l=0.5\xi$, $\xi$ and $2\xi$ and
calculate dependencies $\delta F_{thr}(I)$ at different $\alpha$.
Results of our calculations for length $l=0.5\xi$ are shown in
figure~\ref{Fig:vartc} where we also present fitting expression
$\delta F_{thr}=\delta F_{thr}(0)(1-I/I_c)^b$. For bridge with
critical current $I_c=0.95I_{dep}$ ($\alpha=0.6$) we have
$b\approx 1.36$, for bridge with $I_c=0.74I_{dep}$
($\alpha=-0.55$) $b\approx 1.45$, and bridge with
$I_c=0.66I_{dep}$ ($\alpha=-1.05$) is well fitted by $b\approx
1.5=3/2$ typical for short bridge \cite{JETPL-2016} and Josephson
junction \cite{PRB-1974}. Besides we find that in all cases
$\delta F_{thr}(0) \simeq \hbar I_c/e$ (see inset in figure
\ref{Fig:vartc}) which is typical for a Josephson junction and
resembles result found in framework of GL model both in limiting
cases of long $L\gg \xi$ and short $L \ll \xi$ bridges.

\begin{figure}[hbt]
 \begin{center}
\includegraphics[width=1.0\linewidth]{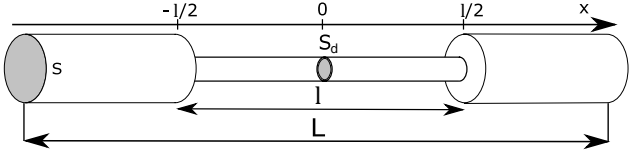}
 \caption{\label{Fig:wire}
Superconducting bridge with the area of cross–section S and length
L, containing constriction with length l and cross–section $S_d$.}
 \end{center}
\end{figure}

The second type of defect models the inhomogeneity of
cross-section area of the bridge. We assume that there is region
with the cross-section area $S_d<S$ and length $l$ in the center
of the bridge (see figure~\ref{Fig:wire}. To describe such a
constriction, the boundary condition ~\eref{bridge-bound} is
supplemented by conditions, similar to the conditions
from~\cite{JETPL-2016} \numparts
\begin{eqnarray}
  \label{def-bridge-bound-dif}
   \frac{df^L}{dx}\bigg|_{-\frac{l}{2}}=\frac{S}{S_{d}}\frac{df^{C}}{dx}\bigg|_{-\frac{l}{2}}, \frac{S}{S_{d}}\frac{df^{C}}{dx}\bigg|_{\frac{l}{2}}=\frac{df^{R}}{dx}\bigg|_{\frac{l}{2}},
  \\
      \label{def-bridge-bound-f}
       f^L\big|_{-\frac{l}{2}}=f^C\big|_{-\frac{l}{2}}=f^C\big|_{\frac{l}{2}}=f^R\big|_{\frac{l}{2}},
\end{eqnarray}
\endnumparts
where $f^{L}, f^{C}, f^{R}$ are the magnitudes of the order
parameter to the left of the defect, in the defect and to the
right of the defect, respectively. The
condition~\eref{def-bridge-bound-dif} appears from the variation
of the Ginzburg--Landau functional for the superconductor with the
cross-section depending on $x$ (which is responsible for the
appearance of the derivative $d/dx(S(x)df/dx)$). Here $S/S_{d}$ is
not the actual ratio of areas of cross-sections but it is a
reference parameter characterizing a change in the derivative of
the function $f$ in $x$ direction at the transition through the
bridge--defect interface.

\begin{figure}[hbt]
 \begin{center}
\includegraphics[width=1.0\linewidth]{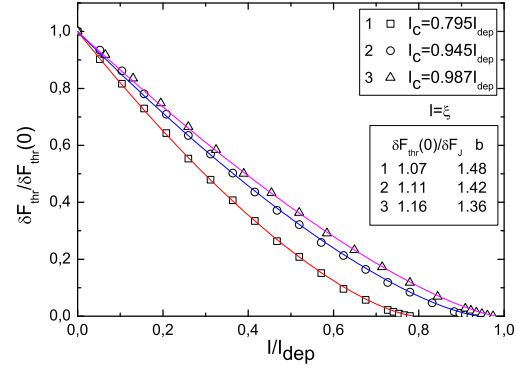}
 \caption{\label{Fig:vars}
Dependence of the energy of threshold fluctuation on current for
bridge with different constrictions (the length of constriction is
fixed: $l=\xi$). Fitting functions $\delta F_{thr}(0)(1-I/I_c)^b$
are shown by the solid lines, the parameters $\delta F_{thr}(0)$
and $b$ are shown in inset. Here $\delta F_J=\hbar I_c/e$.}
 \end{center}
\end{figure}

Calculated dependencies $\delta F_{thr}(I)$ for the constriction
with $l= \xi$ and different cross-sections $S_d$ are shown in the
figure~\ref{Fig:vars} together with fitting expressions $\delta
F_{thr}=\delta F_{thr}(0)(1-I/I_c)^b$. In case of bridge with
$S_d=0.9S$ critical current $I_c=0.987I_{dep}$ and $b\approx
1.36$, for bridge with $S_d=0.76S$ $I_c=0.945I_{dep}$ ($b\approx
1.42$) and for bridge with $S_d=0.5S$ $I_c=0.795I_{dep}$ and
$b\approx 1.48$. This result demonstrate that even small variation
of cross-section area can significantly change dependence $\delta
F_{thr}(I)$ and both power--$5/4$ law and power--$3/2$ law are not
suitable to fit the current dependence of $\delta F_{thr}$. As in
case of local variation $T_c$ even relatively 'weak' constriction
'provides' power law $3/2$ and $\delta F_{thr}(0) \simeq \hbar
I_c/e$ (see inset in figure~\ref{Fig:vars}).

Very similar results could be obtained if in the bridge there is
local variation of mean path length $\ell$ (third type of defect).
In principle, to calculate $\delta F_{thr}(I)$ one can use
analytical results for distribution of $f$ and phase along the
superconducting bridge from \cite{Baratoff} but we use numerical
procedure because dependence $f$ on coordinate is expressed via
special functions. We find that when $\ell$ is five times smaller
in the region with length $l=0.5 \xi$ dependence $\delta
F_{thr}(I) \simeq 1.06 \delta F_J(1-I/I_c)^{3/2}$ with $I_c \simeq
0.73 I_{dep}$.

Change of the exponent for considered types of defect from $5/4$
to $3/2$ could be understood in the following way. In a long
defectless bridge length of the critical nucleus (the region with
suppressed $\Delta$) diverges as $I\rightarrow I_{dep}$
~\cite{PhysRev-164}, while in a bridge with defect its length is
restricted by length of defect plus $\sim 2\xi$ (when $I_c
\lesssim 0.7 I_{dep}$). It resembles the situation with a short
superconducting bridge, which behaves like a Josephson junction
and has $\delta F_{thr}\propto(1-I/I_c)^{3/2}$.

\section{The energy of threshold fluctuation at arbitrary temperature}

Obtained in section II results are valid near critical temperature
$T_c$, since they are based on the Ginzburg--Landau model. Below
we show on the example of defectless bridges that the dependencies
$\delta F_{thr}=\delta F_{thr}(0)(1-I/I_c)^b$ ($b=5/4,3/2$) and
$\delta F_{thr}(0)\propto I_c$ are valid at $T\ll T_c$ too.

At first, we consider, similar to Zharov et al.~\cite{PRL-2007},
the case of a long clean one-dimensional superconducting bridge
($\ell\gg\xi_0$, with $\xi_0=\hbar v_F/\pi\Delta_0$ is the
coherence length in clean limit at $T=0$) containing only one
conduction channel. To find the saddle state in that case we use
the one-dimensional Eilenberger equations for the normal and
anomalous Green's functions, $g(x,\omega_n,v_F)$ and
$f(x,\omega_n,v_F)$ respectively
\begin{eqnarray}
\nonumber
  \hbar v_F \frac{dg}{dx}+\Delta f^{+}-\Delta^{*}f=0,
  \\
  \label{eilenberger}
   -\hbar v_F \frac{df}{dx}-2\omega_nf+2\Delta g=0,
  \\
  \nonumber
  \hbar v_F \frac{df^{+}}{dx}-2\omega_nf^{+}+2\Delta^{*} g=0,
\end{eqnarray}
where $v_F$ is the Fermi velocity, $\omega_n=2\pi k_BT(n+1/2)$ is
the Matsubara frequency. The Green's functions obey the
normalization condition $g^2+ff^{+}=1$. These equations are
completed with the self--consistency equation for the order
parameter $\Delta$
\begin{eqnarray}
 \label{eilenberger-self-cons}
 \frac{\Delta (x)}{\lambda}=\pi N_0 k_BT
 \sum_{\omega_n}\frac{1}{2}[f(x,\omega_n,v_F)+
\\
\nonumber
 +f(x,\omega_n,-v_F)],
\end{eqnarray}
and the expression for the supercurrent density
\begin{eqnarray}
\label{eilenberger-supercurrent}
 j=-2\pi \imath e N_0
 k_BT\sum_{\omega_n}\frac{v_F}{2}[g(x,\omega_n,v_F)-
 \\
  \nonumber
 -g(x,\omega_n,-v_F)].
\end{eqnarray}
Here $\lambda$ is the coupling constant, $N_0$ is the density of
states on the Fermi level. The summation is going over all
Matsubara frequencies.

Following~\cite{PRL-2007}, we seek the solution in the form of
plane waves $\Delta,f\propto e^{\imath kx}$ with complex
amplitudes and solve \eref{eilenberger}. To calculate the energy
of threshold fluctuation, we use the expression
~\eref{fluct-energy} derived by Eilenberger in his work
~\cite{Eilenberger}. Using this expression and saddle-point
solution of~\eref{eilenberger}, one can calculate the energy of
threshold fluctuation
\begin{eqnarray}
 \nonumber
  \delta F_{thr}=SN_0\pi k_BT\hbar v_FRe\sum_{\omega_n}\left(\ln\frac{a_{+}}{a_{-}}-\frac{2\Delta_{R0}}{\sqrt{{\omega'_n}^2+\Delta^2_0}}\right)
  \\
   \label{eqlenberger-en-fluc}
   -\frac{\hbar}{e}I\arctan\frac{\Delta_{R0}}{|\Delta_{I}|}.
\end{eqnarray}
Here
$a_{\pm}=\Delta^2_0-i\Delta_I\omega'_n\pm\Delta_{R0}\sqrt{{\omega'_n}^2+\Delta^2_0}$,
${\omega'}_n=\omega_n+i\hbar k/2$, $\Delta_0$ is the absolute
value of the complex amplitude of the order parameter,
$\Delta_{R0}$ and $\Delta_I$ are the real and imaginary parts of
the complex amplitude, which is determined by the equations
\begin{eqnarray}
 \label{order-abs}
 \pi
 k_BT\sum_{\omega_n}\left(Re\frac{1}{\sqrt{{\omega'_n}^2+\Delta^2_0}}-\frac{1}{|\omega_n|}\right)=\ln
 T,
 \\
 \label{order-im}
  \sum_{\omega_n}Im\left[\frac{1}{\omega'_n+i\Delta_I}\frac{1}{\sqrt{{\omega'_n}^2+\Delta^2_0}}\right]=0.
\end{eqnarray}

\begin{figure}[hbt]
 \begin{center}
\includegraphics[width=1.0\linewidth]{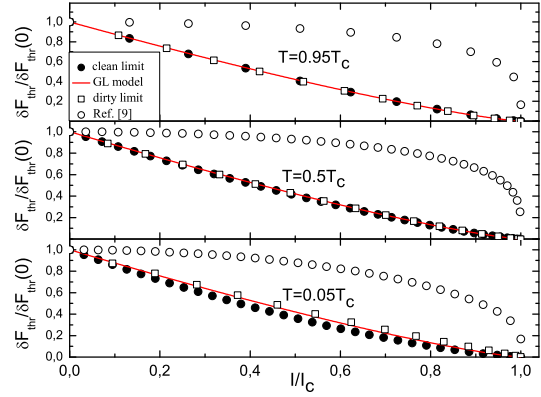}
 \caption{\label{Fig:tarbitr}
Dependence of the energy of threshold fluctuation on current for a
long bridge ($L \gg \xi(T)$) at different temperatures in the
clean and dirty limits. We compare them with the results following
from GL model (solid lines - equation~\eref{df-LA})
and~\cite{PRL-2007} (white circles - equation
\eref{eqlenberger-en-fluc} without last term).}
 \end{center}
\end{figure}

In the work~\cite{PRL-2007} the
expression~\eref{eqlenberger-en-fluc} does not contain last term,
which includes the work performed by the current source on the
system during the transition of the system from the ground state
into the saddle--point state. The comparison of our results with
the results of~\cite{PRL-2007} and the LA theory is shown in the
figure~\ref{Fig:tarbitr}. It is seen that accounting of this term
significantly changes the dependence $\delta F_{thr}(I)$ and
brings it to the form that is similar to~\eref{df-LA} in wide
temperature range below the critical temperature (only at
$T/T_c=0.05$ there is noticeable deviation from the power--$5/4$).

Than we consider the case of dirty superconducting bridge
($l\ll\xi_0$). To calculate the energy of saddle--point state, we
use the Usadel equation~\cite{Usadel} for the normal
$g(\omega_n,x)$ and anomalous $f(\omega_n,x)$ Green's functions in
standard parametrization ~\cite{JLTP-1981}
\begin{eqnarray}
 \nonumber
  g(\omega_n,x)=\cos\theta (\omega_n,x),
  \\
  \label{theta-param}
   f(\omega_n,x)=\sin\theta (\omega_n,x) e^{\imath\chi (x)},
\end{eqnarray}
where $\theta$ and $\chi$ are real functions. With that
parametrization the Usadel equation reads as
\begin{equation}
 \label{usadel}
 \frac{\hbar D}{2}\frac{d^2\theta}{dx^2}-\left(\omega_n+\frac{D}{2\hbar}q_s^2\cos\theta\right)\sin\theta+\Delta\cos\theta=0,
\end{equation}
while the self--consistency equation and the expression for the
supercurrent density takes the form
\begin{eqnarray}
 \label{usadel-self-const}
   \Delta\ln\frac{T}{T_c}=2\pi
   k_BT\sum_{\omega_n>0}\left(\sin\theta-\frac{\Delta}{\omega_n}\right),
   \\
    \label{usadel-supercurrent}
    j=4eN_0D\pi T\frac{q_s}{\hbar}\sum_{\omega_n>0}\sin^2\theta.
\end{eqnarray}
Here $D$ is the diffusion coefficient, $q_s=\hbar(d\chi/dx)$ is
the superfluid momentum. The free energy in~\eref{fluct-energy}
can be written as
\begin{eqnarray}
 \nonumber
 F=2\pi N_0k_BTS\sum_{\omega_n>0}\int dx\left\{\frac{\hbar D}{2}\left[\left(\frac{d\theta}{dx}\right)^2\right.\right.
 \\
  \label{usadel-en-fluc}
  \left.\left.+\left(\frac{q_s
\sin\theta}{\hbar}\right)^2\right]-2\omega_n(\cos\theta-1)-2\Delta\sin\theta\right.
 \\
 \nonumber
  \left.+\frac{\Delta^2}{\omega_n}\right\}+N_0S\int dx\Delta^2\ln\frac{T}{T_c}.
\end{eqnarray}

\begin{figure}[hbt]
 \begin{center}
\includegraphics[width=1.0\linewidth]{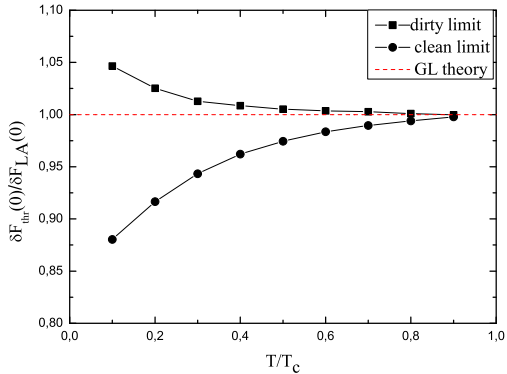}
 \caption{\label{Fig:dfvst}
A temperature dependence of the energy of threshold fluctuation at
zero current for a long bridge ($L \gg \xi(T)$) in the clean and
dirty limits. The energy is measured in units of $\delta
F_{LA}(0)=\sqrt{6} I_{dep}\hbar /2e$.}
 \end{center}
\end{figure}

The equations (\ref{usadel} -- \ref{usadel-supercurrent}) are
numerically solved for a long bridge using the Newton's method
with the boundary conditions $\theta=\theta_{\infty}$ at $x=\pm 15
\xi_{T_c}$ ($\xi_{T_c}=\sqrt{\hbar D/k_BT_c}$), where
$\theta_{\infty}$ is the solution of the uniform Usadel equation
\begin{equation}
 \label{usadel-hom}
  -\left(\omega_n+\frac{D}{2\hbar}q_s^2\cos\theta_{\infty}\right)\sin\theta_{\infty}+\Delta_{\infty}\cos\theta_{\infty}=0.
\end{equation}
Search of the saddle state is performed in a similar way as we do
on the basis of the GL theory with the only difference that we fix
the ratio $\sin\theta (0)/\sin\theta_{\infty}$ at $x=0$ instead of
the magnitude of the order parameter. The dependence $\delta
F_{thr}(I)$ is shown in figure~\ref{Fig:tarbitr}. It can be seen
that for dirty long bridge the current dependence of $\delta
F_{thr}$ remains close to the dependence described by Eq. (1).
Besides $\delta F_{thr}(0) \simeq \delta F_{LA}(0)$ (see
figure~\ref{Fig:dfvst}) in broad range of temperatures below $T_c$
and if one uses for $I_{dep}(T)$ result following from microscopic
calculations and not the Ginzburg-Landau depairing current. In the
clean limit the deviation is stronger, reaching about $15\%$ for
$\delta F_{thr}(0)$ as $T\rightarrow 0$~\cite{PRL-2007}.

Alongside the case of long bridges, we also study short bridges
($L\ll\xi(T)$) in dirty limit. In this case we can neglect
nongradient terms in \eref{usadel} inside the bridge like it was
done by Kulik and Omelyanchuk \cite{JETPL-1975} and we obtain the
following equation
\begin{equation}
 \label{usadel-short}
  \frac{\hbar
  D}{2}\frac{d^2\theta ^{C}}{dx^2}-\frac{D}{2\hbar}q_s^2\cos\theta ^{C}\sin\theta=0,
\end{equation}
where $\theta ^{C}$ defines $\theta$ inside the bridge. In the
work~\cite{JETPL-1975} the solution of this equation was found
together with current-phase relation $I(\phi)$
\begin{equation}
 \label{cpr}
 I(\phi)=\frac{4\pi
  k_BT}{eR_N}\sum_{\omega_n>0}\frac{\Delta_{\infty}\cos\frac{\phi}{2}}{\delta}\arctan\frac{\Delta_{\infty}\sin\frac{\phi}{2}}{\delta},
\end{equation}
where $\delta=\sqrt{(\Delta_{\infty}\cos\phi /2)^2+{\omega_n}^2}$
and $\phi$ is the phase difference across  the bridge.
In~\eref{cpr} for each current there are two values of $\phi$
corresponding to two different states - the smaller $\phi$
corresponds to the ground state, and the larger $\phi$ corresponds
to the saddle state. The strategy to find $\delta F_{thr}$ is
following - for fixed current we find two values of $\phi$, than
with these $\phi$ we use analytical solution from
\cite{JETPL-1975} for $\theta^C$ while for $\theta$ outside the
bridge we numerically solve equations~\eref{usadel} and
~\eref{usadel-self-const}, neglecting by the pair breaking effect
of the current/supervelocity in the banks (which is applicable
when cross-section of banks $S_{pad}\gg S$). Solutions in the
bridge and in the banks are matched by using the boundary
conditions
\begin{eqnarray}
  \nonumber
   \frac{d\theta^{L,R}}{dx}\bigg|_{\pm\frac{L}{2}}=\frac{S}{S_{pad}}\frac{d\theta^{C}}{dx}\bigg|_{\pm\frac{L}{2}}
   \\
   \label{pad-bridge-bound-dif-theta}
   =\frac{S}{S_{pad}}\frac{2}{L}\frac{\omega_n \sin\frac{\phi}{2}\arctan\frac{\Delta_{\infty}\sin\frac{\phi}{2}}{\delta}}{\delta},
    \\
      \label{pad-bridge-bound-theta}
       \theta^L\big|_{-\frac{L}{2}}=\theta^C\big|_{-\frac{L}{2}}=\theta^C\big|_{\frac{L}{2}}=\theta^R\big|_{\frac{L}{2}},
       \\
        \label{sys-bound-theta}
        \theta^L\big|_{-\frac{L_{sys}}{2}}=\theta^R\big|_{\frac{L_{sys}}{2}}=\theta_{\infty},
\end{eqnarray}
where $\theta^{L}, \theta^{R}$ are the functions $\theta$ in the
left bank and right bank, respectively. Here $L_{sys}=40
\xi_{T_c}+L$ is length of modelled system, including the bridge
(with length L) and the banks with cross-section $S_{pad}$ and
length $(L_{sys}-L)/2$ which are contacted with much wider banks
where $\theta$ is equal to its value at given temperature and zero
current. Above conditions appear from the conservation law for
spectral currents~\cite{Zaitsev} and is similar to the boundary
conditions (8).

\begin{figure}[hbt]
 \begin{center}
\includegraphics[width=1.0\linewidth]{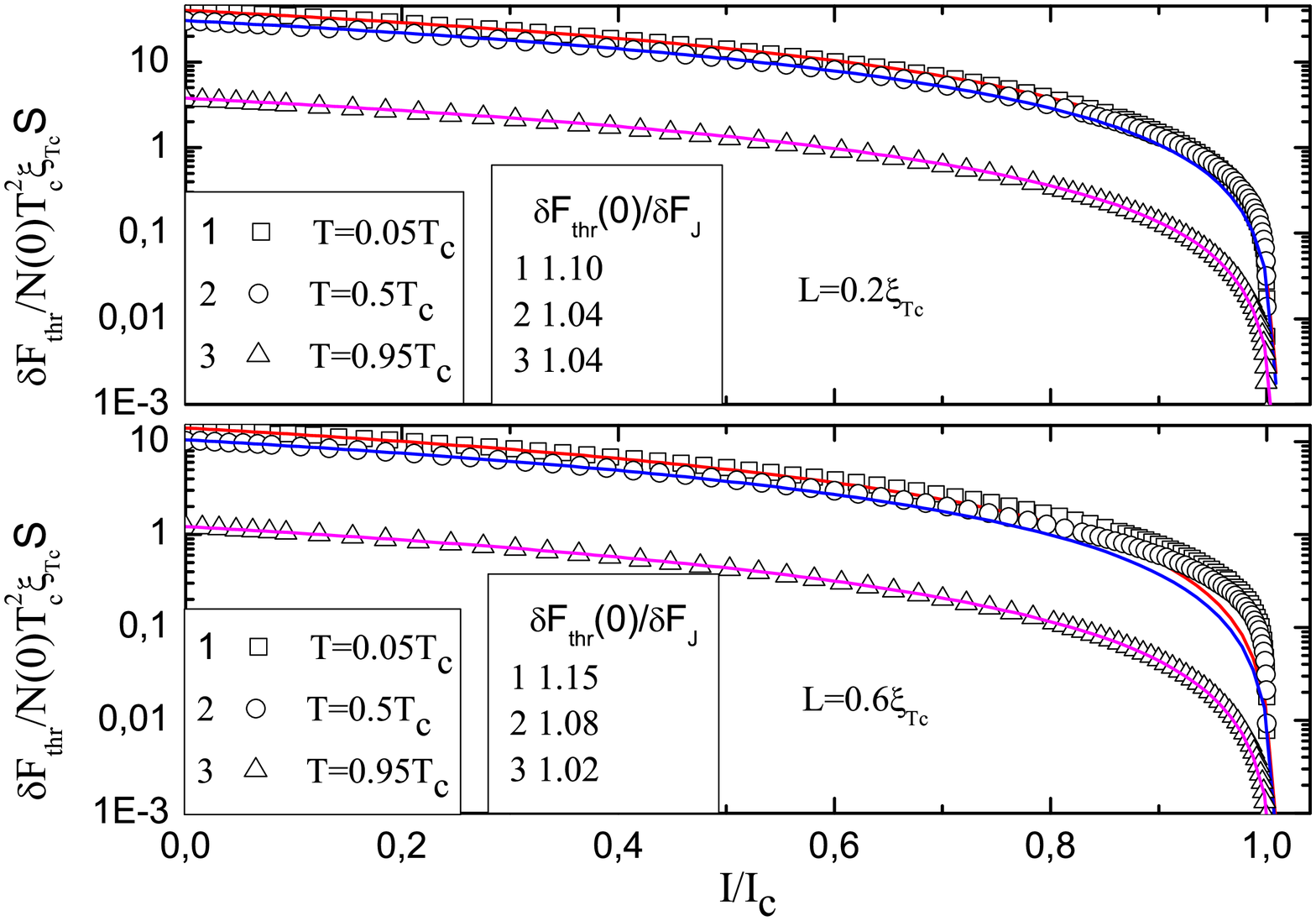}
 \caption{\label{Fig:dfshort}
A current dependence of the energy of threshold fluctuation for
short bridges ($L=0.2\xi_{T_c}$ and $L=0.6\xi_{T_c}$) at different
temperatures. The functions $\delta F_{thr}(I=0) (1-I/I_c)^{3/2}$
are shown by the solid lines. Here $\delta F_J=\hbar I_c/e$.}
 \end{center}
\end{figure}

Calculated $\delta F_{thr}(I)$ are shown in
figure~\ref{Fig:dfshort}. For $L\ll\xi(T)$ ($L=0.2\xi_{T_c}$) the
power--$3/2$ law is approximately valid at all temperatures (note
noticeable difference at $I \gtrsim 0.8 I_c$ for $T=0.5 T_c$ and
$T=0.05 T_c$) while for bridge with $L=0.6\xi_{T_c}$ the condition
$L\ll\xi(T)$ is not applicable at low temperatures, which leads to
stronger deviation from the power--$3/2$ law in wide range of
currents near $I_c$. Note, that $\delta F_{thr}(0) \simeq \hbar
I_c/e$ (see inset in ~\ref{Fig:dfshort}) with the largest
deviation at low temperatures.

\begin{figure}[hbt]
 \begin{center}
\includegraphics[width=1.0\linewidth]{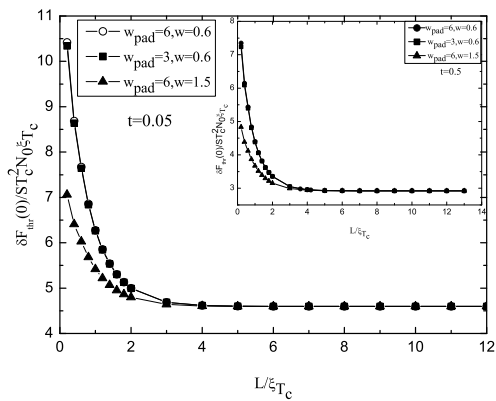}
 \caption{\label{Fig:df2d}
Energy of threshold fluctuation versus the length of the bridge at
zero current at different widths of the bridge and banks. On the
main graph results for the temperature $T=0.05T_c$ are shown,
while in the inset results for $T=0.5T_c$ are present.}
 \end{center}
\end{figure}

And finally, in dirty limit we find how $\delta F_{thr}(0)$
depends on the length of the bridge. Earlier, in work
~\cite{JETPL-2016} we claimed that dependence $\delta
F_{thr}(0,L)$ may have a minimum at $L \simeq 2-3 \xi(T)$ at
proper choice of widths of banks and bridge. We carried
calculations (to determine the saddle--state, the condition
$\theta (x=0, y)=0$ is added) using two-dimensional Usadel
equation in the the same geometry as in ~\cite{JETPL-2016} (see
figure 4 there) and the same geometrical parameters but we did not
find a minimum (see figure~\ref{Fig:df2d}). Instead $\delta
F_{thr}(0)$ monotonically increases when $L$ decreases following
increase of $I_c$. This result force us to check our calculations
made in framework of GL model ~\cite{JETPL-2016} and we find that
this result is an artefact of used grid approximation. With proper
grid we confirm absence of minimum in dependence $\delta
F_{thr}(L)$ in GL model too.

\section{Discussion}

We demonstrate, that functional dependence of energy of threshold
fluctuation (perturbation) on current following from
Ginzburg-Landau model stays valid at temperatures well below $T_c$
both in dirty and clean limits if one uses actual critical
(depairing) current but not the Ginzburg-Landau depairing current.
This result gives us the hope that the strong effect of even
relatively 'weak' defect (which does not strongly suppress
critical current of the bridge and provide $I_c \simeq I_{dep}$)
on dependence $\delta F_{thr}(I)$ that was found at $T\sim T_c$ is
temperature independent and could be applicable at low
temperatures, too.

Our results could be used for qualitative explanation of the
dependence $\delta F_{thr}(I)\sim (1-I/I_c)^{3/2}$ found in
works~\cite{PRL-2011,Aref} for long bridges/wires by presence  of
intrinsic defects in their samples. Unfortunately we are not able
to make quantitative comparison due to lack of important
parameters (resistivity and diffusion coefficient of the
bridges/wires, their width and thickness) which are needed to see
how far the actual critical current of the bridge is from the
depairing current. Alternative explanation of that experiments is
based on the model of the bridge/wire as chain of weakly
connected, via Josephson coupling, granules \cite{Khlebnikov}
which naturally leads to power --$3/2$ but it is not clear how
this model could be applicable to works ~\cite{PRL-2011,Aref}.

\section{Conclusion}

We calculate the energy of threshold fluctuation which switches
the current-carrying superconducting bridge to resistive state. We
make calculations at arbitrary temperature, different length of
the bridge and in presence of defects connected with local
variation $T_c$, mean path length $\ell$ or cross-section of the
superconductor. It is found that the presence of defect has strong
influence on the form of current dependence of the energy of
threshold fluctuation, changing it from $\delta F(I) \simeq
(1-I/I_{dep})^{5/4}$ valid for long defectless bridge to $\delta
F(I) \simeq (1-I/I_c)^{3/2}$ which is typical for short bridge and
Josephson junction. Additionally, using microscopic theory we show
that the results, obtained on the basis of Ginzburg--Landau
theory, stay valid at temperatures significantly below $T_c$, if
one uses proper temperature dependent critical (depairing)
current.

\ack

The study is supported by the Russian Foundation for Basic
Research (grant No 15-42-02365).

\end{document}